\documentclass[aps,prd,superscriptaddress,twocolumn]{revtex4-1}
\usepackage{amsmath}
\usepackage{bm}
\usepackage{times}
\usepackage{braket}
\usepackage{color,graphicx}

\begin{document}
\title{Scalar leptoquark in $SU(5)$}

\author{Ilja Dor\v sner} \email[Electronic address:]{ilja.dorsner@ijs.si}
\affiliation{Department of Physics, University of Sarajevo, Zmaja od Bosne 33-35, 71000
  Sarajevo, Bosnia and Herzegovina}

\begin{abstract}
We address the issue of model dependence of partial proton decays due to exchange of a single scalar leptoquark within a minimal viable $SU(5)$ framework. The minimal setup predicts a flavor part of decay widths for $p \to \pi^+  \bar{\nu}$ and $p \to K^+ \bar{\nu}$ processes to depend solely on the known masses and mixing parameters of the quark sector and one extra phase. We accordingly establish an accurate lower limit on the mass of the scalar leptoquark in connection with the relevant experimental constraints on the matter stability. The ratio of decay widths for $p \to \pi^+  \bar{\nu}$ and $p \to K^+ \bar{\nu}$ channels turns out to be phase independent and predicts strong suppression of the former width with respect to the latter one. Our results offer a possibility to test the minimal scenario if and when proton decay is observed.
\end{abstract}
\maketitle


\section{Introduction}

One of the most appealing features of the Georgi-Glashow model of unification~\cite{Georgi:1974sy} is a predictive nature of the original setup. It is this very property that helped conclusively rule out the model. One problematic issue of the minimal framework is a conflict between predicted and observed fermion masses and mixing parameters. Another one is an inability of the model to accommodate unification of gauge couplings.

The predictive nature of the original Georgi-Glashow proposal is also reflected in implications it yields for the matter stability. The model predicts that all partial proton decay widths exhibit dependence on a single unitary matrix~\cite{Mohapatra:1979yj} that can be identified with a Cabibbo-Kobayashi-Maskawa (CKM) mixing matrix, albeit one that contains additional diagonal unitary transformations. This result, however, is valid under an explicit assumption that neutrinos are massless particles thus allowing one to neglect the leptonic mixing altogether. Moreover, the fact that the down quark and charged lepton mass matrices are transpose of each other in the original model automatically eliminates uncertainty related to the transition from an arbitrary flavor basis to the mass eigenstate basis via appropriate unitary transformations on the matter fields. This property, unfortunately, generates erroneous predictions for the masses and mixing parameters of the quark sector.

The departure from minimality through introduction of a single additional Higgs field~\cite{Georgi:1979df} to address experimentally observed difference between masses in the down quark and charged lepton sectors severely affects the predictive features of the original model. In particular, the flavor part of the partial decay widths for proton decay through scalar exchanges becomes dependent on a completely unknown mismatch between the relevant unitary transformations in the quark and lepton sectors~\cite{Dorsner:2012nq}. To make matters worse, the number of available proton decay channels is smaller than the number of unknown parameters entering associated partial decay widths. It thus seems that even the simplest of departures from minimality spoils the predictive nature of the minimal unification scheme. 

The question we want to pursue in view of all this is whether it is possible to simultaneously fix fermion mass relations while preserving some of the flavor independence for the proton decay predictions of the minimal model. We opt for the most minimal viable extension of the original $SU(5)$ model that accordingly has only one proton mediating scalar field. 


\section{The Setup}

The charged fermion sector of $SU(5)$, in a context of a minimal viable extension, requires use of higher-dimensional operators~\cite{Ellis:1979fg}. The most minimal set of operators that needs to be included in order to have realistic charged fermion masses is
\begin{eqnarray}
\label{eq1}
\epsilon_{\alpha \beta \gamma \delta \eta}  Y^{10}_{ij}  \bm{10}_i^{\alpha \beta}  
\bm{10}_j^{\gamma \delta}  \bm{5}^\eta,\\
\label{eq2}
Y^{\overline{5}}_{1\,ij}  \bm{10}^{\alpha \beta}_i   \overline{\bm{5}}_{\beta j}  \bm{5}^*_\alpha,\\ 
\label{eq3}
Y^{\overline{5}}_{2\,ij}   \bm{10}^{\alpha \beta}_i  \frac{\bm{24}^\gamma_\beta}{\Lambda}  \overline{\bm{5}}_{\gamma j} \ \bm{5}^*_\alpha,
\end{eqnarray}
where $\Lambda$ is the cut-off scale and $i,j(=1,2,3)$ represent the family indices. The matter fields belong to $\bm{10}_i=\{e^C_i,u^C_i,Q_i\}$ and $ \overline{\bm{5}}_{j}=\{L_j,d^C_j\}$, where $Q_i=(u_i
\quad d_i)^T$ and $L_j=(\nu_j \quad e_j)^T$~\cite{Georgi:1974sy}. $Y^{10}$, $Y^{\overline{5}}_{1}$ and $Y^{\overline{5}}_{2}$ are arbitrary complex $3 \times 3$ Yukawa matrices. The contractions in both the $SU(5)$ group space and flavor space are explicitly shown. 

The only scalar representations we include are $\bm{24}$ and $\bm{5}$. The former one breaks $SU(5)$ down to $SU(3) \times SU(2) \times U(1)$ while the latter provides electroweak vacuum expectation value (VEV). We take, in what follows, $|\langle \bm{5}^5 \rangle| = |v|=\sqrt{2}\,246\,\mathrm{GeV}$ and
\begin{equation*}
\langle \bm{24} \rangle \approx \sigma\left(
\begin{array}{ccccc}
2 & 0 &  0 &  0 &  0\\
0 & 2 &  0 &  0 &  0\\
0 & 0 & 2 &  0 &  0\\
0 & 0 &  0 & -3 &  0\\
0 & 0 &  0 &  0 & -3
\end{array}
\right)
\end{equation*}
to be the relevant VEVs. We chose to neglect a possible VEV for electrically neutral component of an $SU(2)$ triplet in $\bm{24}$. This choice does not affect the outcome of our study. We will also demonstrate that our findings will not require one to specify $\Lambda$ and $ \sigma$ in order to generate accurate predictions.

The mass matrices for the charged matter fields, in an obvious notation, come out to be
\begin{eqnarray}
M_U = - \sqrt{2} \left(Y^{10}+Y^{10\,T}\right) v,\\
M_D =   \left(- \frac{1}{2} Y^{\overline{5}}_1  + Y^{\overline{5}}_2 \frac{\sigma}{\Lambda} \right) v^*,\\
M^T_E  =  \left(- \frac{1}{2} Y^{\overline{5}}_1 -\frac{3}{2} Y^{\overline{5}}_2 \frac{\sigma}{\Lambda} \right)v^*.
\end{eqnarray}
The Georgi-Jarlskog type of correction~\cite{Georgi:1979df} is generated by the operator of Eq.~\eqref{eq3} that accordingly breaks unwanted degeneracy between the down quark and charged lepton sectors~\cite{Ellis:1979fg}. This renders the $SU(5)$ setup viable with regard to observed charged fermion masses. We opt not to specify the exact mechanism of the neutrino mass generation as it does not interfere with our results. The lepton mixing matrix is also irrelevant for our study, as we show later. We also do not delve into another important issue---unification of gauge couplings---in our work. Note, however, that two successful descriptions of neutral lepton sector in $SU(5)$ both provide a source for satisfactory unification of gauge couplings~\cite{Dorsner:2005fq,Bajc:2006ia}.

We define the physical basis for the charged fermions through the following set of bi-unitary transformations: $U^T M_U U_C = M_U^{\textrm{diag}}$, $ D^T M_D D_C= M_D^{\textrm{diag}}$ and $E^T M_E E_C = M_E^{\textrm{diag}}$. Here, $U^{\dagger} D \equiv V_{UD}=K_1 V_{CKM}
K_2$, where $K_1$ ($K_2$) is a diagonal matrix containing three
(two) phases. In the neutrino sector we have $N^T M_N N = M_N^{\textrm{diag}}$ with $E^\dagger
N  \equiv V_{EN}=K_3 V_{PMNS}$. $K_3$ is a diagonal unitary matrix. $V_{CKM}$ ($V_{PMNS}$) is the Cabibbo-Kobayashi-Maskawa (Pontecorvo-Maki-Nakagawa-Sakata) mixing matrix. 

It is evident from Eq.~\eqref{eq1} that $M_U$ is a symmetric matrix. This allows us to take $U = U_C K_0$, where $K_0$ is a diagonal matrix containing three phases. Of all the unknown phases in $K_a$ ($a=0,1,2,3$) matrices, only one will be featured in our results. It is the phase associated with the $(K_0)_{11}$ element. We furthermore take $v(=|v|)$ to be real to simplify our notation. With these preliminaries out of the way we are ready to investigate the couplings of the proton decay mediating scalar in the physical basis. This we do in the next section.


\section{Proton Decay Leptoquark}

There is only one color triplet scalar that contributes to proton decay. It resides in the five-dimensional scalar representation. The leading order baryon number violating dimension-six operators due to its exchanges are~\cite{Nath:2006ut}
\begin{eqnarray*}
\textit{O}(d_{j}, d_{k}, \nu_i) & = & a(d_{j}, d_{k}, \nu_i) \ u^T \ L \ C^{-1} \ d_{j} \ d_{k}^T \ L \ C^{-1}\ \nu_i, \\
\textit{O}(d_{j}, d_{k}^C, \nu_i) & = & a(d_{j}, d_{k}^C, \nu_i) \ {d^C_{k}}^{\dagger} \ L \ C^{-1} \ {u^C}^* 
\ d_{j}^T \ L \ C^{-1}\ \nu_i, 
\end{eqnarray*}
where $i(=1,2,3)$ and $j,k(=1,2)$ ($j+k<4$) represent flavor indices while $L=(1- \gamma_5)/2$. Our notation is such that, for example, $d_1$ ($d_2$) stands for the $d$ ($s$) quark. The color contraction via the Levi-Civit\` a tensor in the $SU(3)$ space is understood. 

$\textit{O}(d_{j}, d_{k}, \nu_i)$ and $\textit{O}(d_{j}, d_{k}^C, \nu_i)$ operators contribute exclusively to proton decay processes with anti-neutrinos in the final state. The relevant coefficients for the $p \to \pi^+ \bar \nu$ ($p \rightarrow K^+ \bar{\nu}$) decay have $j+k=2$ ($j+k=3$)~\cite{Nath:2006ut}. For example, the relevant decay width for $p \to \pi^+ \bar \nu$ is
\begin{equation}
\label{gamma_1}
\Gamma_{p \to \pi^+ \bar \nu} =\sum_{i} 
C_{\pi^+} |\alpha \, a(d_{1}, d_{1}^C, \nu_i) + \beta \, a(d_{1}, d_{1}, \nu_i)|^2,
\end{equation}
where $\alpha$ and $\beta$ are the so-called nucleon matrix elements. We also introduce flavor independent constant
\begin{equation}
C_{\pi^+} = \frac{(m_p^2-m_{\pi^+}^2)^2}{32\pi f_\pi^2 m_p^3} (1+D+F)^2,
\end{equation}
where $F$ and $D$ are extracted from relevant linear combinations that enter the form factors in semileptonic hyperon decays and nucleon axial charge~\cite{Claudson:1981gh, Aoki:2008ku}. We take $f_\pi =130$\,MeV, $D = 0.80$, $F = 0.47$ and $\alpha=-\beta =-0.0112\,\textrm{GeV}^3$~\cite{Aoki:2008ku} when needed. The sum is inserted to take into account the fact that the proton decay experiments do not distinguish the neutrino flavor.

The minimal viable set of operators, as given in Eqs.~\eqref{eq1}--\eqref{eq3}, yields  
\begin{eqnarray}
a(d_{j}, d_{k}, \nu_i)&=& \frac{2}{m^2 v^2} (M_U^{\textrm{diag}} K_0 V_{UD})_{1 j}  (D^T M_D N)_{k i},\,\\
a (d_{j}, d_{k}^C, \nu_i) &=& -\frac{2}{m^2 v^2}  (V_{UD} M_D^{\textrm{diag}})_{1 k}  
(D^T M_D N)_{j i},\,
\end{eqnarray}
where $m$ represents the mass of the exchanged color triplet scalar. The form of $a(d_{j}, d_{k}, \nu_i)$ and $a (d_{j}, d_{k}^C, \nu_i)$ coefficients is particularly simple because the couplings of the color triplet in $\bm{5}$ to matter fields are aligned with the corresponding couplings of the $SU(2)$ doublet in $\bm{5}$ even though we include one higher-dimensional operator to correct fermion masses. The summation over neutrino species conveniently eliminates remaining flavor dependence from the decay widths for $p \to \pi^+ \bar \nu$  and $p \to K^+ \bar \nu$ channels through the following identity
\begin{equation}
\label{sum}
\sum_{i=1,2,3} (D^T M_D N)_{\alpha i}(D^T M_D N)^*_{\beta i}= (M_D^{\textrm{diag}\,2})_{\alpha \beta}.
\end{equation}
Clearly, the lepton mixing matrix does not affect proton decay signatures through scalar exchange. It is also clear that the $p \to \pi^+ \bar \nu$ decay rate is significantly suppressed compared to the $p \to K^+ \bar \nu$ one. The suppression factor, as inferred from Eq.~\eqref{sum}, is proportional to $(m_d/m_s)^2$. 

For the decay widths for $p \to \pi^+ \bar \nu$ and $p \to K^+ \bar \nu$ channels we find
\begin{eqnarray*}
\Gamma_{p \to \pi^+ \bar \nu} &=& C_{\pi^+} \, A\,(m_u^2+m_d^2+2 m_u m_d \cos \phi) m_d^2,\\
\Gamma_{p \to K^+ \bar \nu} &\approx& C_{K^+} \, A\,(m_u^2+m_d^2+2 m_u m_d \cos \phi) m_s^2 ,
\end{eqnarray*}
where we neglect terms suppressed by either $(m_d/m_s)^2$ or $|(V_{UD})_{12}|^2$ in the expression for $\Gamma_{p \rightarrow K^+ \bar{\nu}}$. Here, $A=4 |\alpha|^2 |(V_{UD})_{11}|^2/v^4$, $e^{i \phi}=(K_0)_{11}$ and we introduce 
\begin{equation}
C_{K^+} = \frac{(m_p^2-m_{K^+}^2)^2}{32\pi f_\pi^2 m_p^3} \left(1+\frac{m_p}{3 m_\Lambda}(D+3 F) \right)^2.
\end{equation}
After we insert all low-energy parameters we find 
\begin{equation}
\Gamma_{p \rightarrow \pi^+ \bar{\nu}}/\Gamma_{p \rightarrow K^+ \bar{\nu}}=10^{-2}.
\end{equation}
We take values of quark and lepton masses at $M_Z$, as given in Ref.~\cite{Dorsner:2006hw}. The CKM angles are taken from Ref.~\cite{Nakamura:2010zzi}. Note that our result for the decay widths ratio does not depend on the exact value of the nucleon matrix element and is rather insensitive to the running of relevant operators, quark masses and corresponding mixing parameters. It also does not contain dependence on the extra phase although $\phi$ enters individual widths. This ratio can thus be considered as a firm prediction within the framework of the $SU(5)$ theory with the most minimal set of operators that renders the scenario phenomenologically viable. This, again, is rather an unexpected result as any departure from the Georgi-Glashow model usually affects the predictive nature of the original $SU(5)$ setup. It is gratifying to see that the minimal viable $SU(5)$ scenario manages to preserve some of the most salient features~\cite{Mohapatra:1979yj} of the original proposal~\cite{Georgi:1974sy}.

We can also establish a lower limit on the color triplet scalar mass due to experimental constraints on proton stability. The explicit dependence on the unknown phase $\phi$ can be easily accommodated. We use the most current constraint---$\tau_{p \rightarrow K^+ \bar{\nu}}> 4.0 \times 10^{33}$\,years~\cite{Miura:2010zz}---to obtain the following limits on $m$ in the least ($\phi=\pi$) and the most conservative ($\phi=0$) case
\begin{eqnarray}
\label{limit1}
m_{\phi=\pi}&>& 1.3 \times 10^{11} \left(\frac{|\alpha|}{0.0112\,\mathrm{GeV}^3}\right)^{1/2}\,\mathrm{GeV},\\
\label{limit2}
m_{\phi=0}&>& 2.2 \times 10^{11} \left(\frac{|\alpha|}{0.0112\,\mathrm{GeV}^3}\right)^{1/2}\,\mathrm{GeV}.
\end{eqnarray}

We have only one unknown phase entering decay widths for $p \to \pi^+ \bar \nu$  and $p \to K^+ \bar \nu$. This makes the minimal viable $SU(5)$ a perfect candidate to put the idea of a flavor goniometry through proton decay to the test~\cite{DeRujula:1980qc}. The proton decay modes into charged anti-leptons, on the other hand, depend on {\it a priori} unknown unitary rotations in the flavor space. This prevents us from making any firm predictions with regard to those channels. That is the reason we do not present analysis of those partial decays here. This, however, does not invalidate the limit on the triplet mass of Eqs.~\eqref{limit1} and \eqref{limit2} as the the most constraining limit on the scalar leptoquark mass always originates from the $p \rightarrow K^+ \bar{\nu}$ channel in $SU(5)$~\cite{Dorsner:2012nq} framework. In other words, if the scalar exchange dominates over gauge boson exchange with regard to proton decay, a preferred channel is always $p \rightarrow K^+ \bar{\nu}$.


\section{Summary}

We show that the most minimal viable $SU(5)$ setup predicts a flavor part of decay widths for $p \to \pi^+  \bar{\nu}$ and $p \to K^+ \bar{\nu}$ processes due to scalar exchange to depend solely on the known masses and mixing parameters of the quark sector and one extra phase. We accordingly establish an accurate lower limit on the mass of the scalar leptoquark in connection with the current experimental data on proton stability. The bound on the $p \rightarrow K^+ \bar{\nu}$ channel constrains the color triplet mass to be slightly above $10^{11}$\,GeV. We also show that $p \to \pi^+  \bar{\nu}$ decay rate is predicted to be two orders of magnitude below the rate for $p \to K^+ \bar{\nu}$ process. This offers a possibility to test the scenario if and when proton decay is observed. 


\begin{acknowledgments}
I.D.\ acknowledges the SNSF support through the SCOPES project No.\ IZ74Z0\_137346. I.D.\ thanks B.\ Bajc, S.\ Fajfer and N.\ Ko\v snik for discussions and valuable comments.
\end{acknowledgments}


\end{document}